# Weightless Neural Networks (WNNs) for Continuously Trainable Personalized Recommendation Systems

Rafayel Latif
Animo Omnis Labs
London, United Kingdom
rafayel@aolabs.ai

Satwik Behera
Animo Omnis Labs
Cupertino, CA, USA
satwik@aolabs.ai

Ali Al-Ebrahim
Animo Omnis Labs
Berkeley, CA, USA
ali@aolabs.ai

## ABSTRACT

Given that conventional recommenders, while deeply effective, rely on large distributed systems pre-trained on aggregate user data, incorporating new data necessitates large training cycles, making them slow to adapt to real-time user feedback and often lacking transparency in recommendation rationale. We explore the performance of smaller personal models trained on per-user data using weightless neural networks (WNNs), an alternative to neural backpropagation that enable continuous learning by using neural networks as a state machine rather than a system with pretrained weights. We contrast our approach against a classic weighted system, also on a per-user level, and standard collaborative filtering, achieving competitive levels of accuracy on a subset of the MovieLens dataset. We close with a discussion of how weightless systems can be developed to augment centralized systems to achieve higher subjective accuracy through recommenders more directly tunable by end-users.

## CCS CONCEPTS

• **Information systems** → **Recommender system; Users and interactive retrieval.**

## KEYWORDS

Neural networks, Weightless neural networks (WNNs), State machines, Collaborative filtering, Real-time, Reinforcement learning, Personalization

## 1 INTRODUCTION

Industry demand for better recommenders often means that, in production, systems are trained on increasingly larger datasets, a noted trend in machine learning writ large [1]. In a sense, while accuracy has improved, there has been a tradeoff with respect to flexibility– large models require increasingly larger training cycles. Often such batch-training is more temporally dispersed as a result, while also requiring massed user data for optimal performance, inhibiting models from interacting with user feedback in real-time [2]. Advances in this space have involved radically different approaches yet often still involve aggregating user data.

In this paper we explore a novel weightless approach that can utilize even less data, down to the per-user level. As part of this effort we benchmarked a weightless implementation to a comparable per-user implementation using a classic neural network (PyTorch), measuring the average accuracy of the predicted users' sentiment of the movie on a per-user basis, training a new agent for each new user. We also have tested an industry standard approach, collaborative filtering, against the same data though in aggregate, to explore the implications for data drift and the cold start problem.

## Weightless neural networks: in a nutshell

WNNs were largely developed at Imperial College in the 1980s and, like many other machine learning methods, have been overshadowed by the popularity of backpropagation or gradient-descent techniques following the seminal success of AlexNet in 2012 [3]. WNNs have recently seen more popularity as their weightless nature can enable more efficient inference at the edge [4].

Briefly, WNNs function as neural state machines– rather than activating according to a set of stored weights, each neuron for inference uses a table lookup operation [5]. In our particular implementation, each neuron's state can be continuously extended even at inference time, affecting inference to build a learning loop. Many WNN architectures have high memory requirements [4], which we offset by using a smaller dataset that in production can also be trained by real-time user feedback.

The models used in this paper involved a 3-layer WNN architects– input and output layers, and a hidden layer which has a shape equal to that of the input layer [5]. The three layers are connected in a recurrent fashion: input to inner state, inner state to output. Each neuron's lookup table stores a subset of the global information depending on that neuron's connection to others. Each time we query the network, the neurons use CGA [5] to compute either discrimination distance or Hamming distance between novel input and their lookup tables to decide on a binary output (1 or 0). The output of each neuron moves the ensemble into various stable states which approximate the overall input to the training data, and given a learning signal the input-output pair can be added to the training data. This allows for adaptable generalization and continuous or online training, getting us closer to models that can learn users' changing preferences in real-time, giving users better recommendations by giving them better control and tunability over their recommenders, blurring the line between recommendation and active curation.

## Cold start issue

Perhaps the most well-known and difficult challenge in recommender systems is the cold start problem. How can we provide meaningful recommendations when we have little to no prior data about a user or even about the system as a whole? Traditional collaborative filtering models depend heavily on user-item interaction histories. They learn by finding patterns across many users and items, typically by training user and item embeddings. However, when there is insufficient data — for example, with new users (user cold start) or new items (item cold start) [6]— these systems struggle to make accurate predictions, or cannot function at all.

Personalized models offer a compelling alternative. By treating each user as a separate agent and training a model on a per-user basis, preferences can be quickly and independently adapted. Even starting with minimal interaction history (e.g. a handful of likes or dislikes), the perception of increased control –that these models learn your continual feedback– incentivizes users' continued feedback to the system, improving their experience with more training. We collaborated with researchers from the University of Middlesex to build a YouTube recommender chrome extension to gauge the qualitative aspects of these assumptions with user surveys, comparing users' preferences of recommendations received from YouTube's algorithm with those from the individually-trained weightless extension.

## Data drift issue

In current recommender systems, user preferences are often assumed to be static [7]. However, in reality, people's tastes evolve and their interests shift over time. Despite this, many systems struggle to capture and adapt to these dynamic changes effectively. This is particularly problematic in static or batch-trained models that rely heavily on historical data without accounting for recency or context.

Personalized models trained per-user can offer a solution as they can better adapt to individual changes in interest. Additionally, by incorporating contextual features such as time of day, seasonality, or device type as inputs to the model, we can learn more nuanced user behavior patterns. This allows for improved responsive and context-aware recommendations that evolve alongside the user. Also, in a weightless system, the training pairs are stored in the state memory, so they can be selectively deleted to modify recommendation behavior, unlike weighted systems where such credit assignment is not as straightforward and in deep learners is further obscured by hyperparameterization (the blackbox problem). Incorporating richer user data to compensate for aggregate user data also could contribute to solving filter-bubble like effects and the increased polarization of preferences observed in large networks of users [8].

## Quantitative Analysis - Method & Results

As part of this effort, we put together a series of quantitative tests to benchmark WNNs against both a classic weighted PyTorch neural network and a Collaborative Filtering algorithm [8]. We did this via "The Movies Dataset" [9], a dataset of over 26 million reviews with 270k different reviewers. The WNN and PyTorch models were run on a per-user basis, creating a new agent for every user while the collaborative filtering model was run on varying numbers of users but taking the approach of training one model on all the user data. All of the models used the same data encoding method to encode the inputs into binary as shown in the table below:

| Input | Number of binary neurons needed for encoding |
|---|---|
| Genre | 10 |
| Amount of reviews | 10 |
| Average rating | 3 |
| Language | 3 |

This resulted in 26 input neurons, which the models use to predict the rating a movie earned on a scale of 0.5-5 in 0.5 increments (10 output values/classes). The ratings were encoded as 10-dimensional cumulative binary vectors, where a rating of r was represented by setting the first 2r items in a length 10 array of 0s to ones to 1 (e.g., 1.0 → [1, 1, 0, ..., 0], 1.5 → [1, 1, 1, 0, ..., 0]). The models' output values were considered accurate if within +/- 1 of the actual review score (e.g. if the actual rating was 3 out of 5, then the value of 2.5 to 3.5 would be considered correct)

| Reviews Per User | Accuracy (Weighted Network) | Accuracy (WNN) | Accuracy (Collaborative Filtering) |
|---|---|---|---|
| 5 | 0.584 | 0.74 | 0.804 |
| 10 | 0.648 | 0.762 | 0.83 |
| 25 | 0.6656 | 0.7712 | 0.8592 |
| 100 | 0.6672 | 0.7636 | 0.861 |
| 200 | 0.6668 | 0.7737 | 0.8311 |

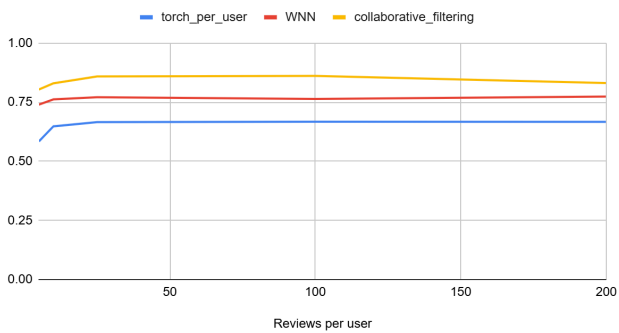

*Notes: all of the tests were run on 250 users, the WNN and TF models were run on a per-user basis while the Collaborative model used a combined approach, taking in all user data for the amount of reviews.*

The table shows the WNNs excel where the number of ratings per-user is lower. For the first row of the table the WNN achieved an accuracy of **74**% while the pytorch model achieved **58.4**% and the collaborative model got **80.4**% showing the potential of the system when we have a low number of training examples. We also see the collaborative model completely failing even though it is trained on all of the users data at once, especially struggling when we have a small amount of user data (250 users). The WNN however are not as affected by the amount of training data which highlights the utility for per user models.

## Challenges with WNNs

While personalization and continuous learning are huge advantages, the nature of WNNs can be problematic. First, the AO Labs agent only accepts binary input and the size of the inputs cannot change after the user has initialized the agent. This at first seems fine - the inputs can be coded into binary (e.g. for genders, male: 00, female: 01 etc). However, what happens if there is an undefined input space - i.e. what if there are practically an infinite amount of genres to choose from? How do we encode them into binary while not overloading the system to the point where the binary encodings are meaningless? In a YouTube video recommender [2], this issue arose when trying to use genres as an input for the agent. There is an undefined amount of genres to choose from, so how is this issue overcome? This can be done with the help of embedding models. Two options arise. One is to convert the input into an embedding via an embedding model, and convert that into a fixed amount of binary digits using a gaussian random projection model (or similar). The issue with this is that the agent would require a min of about 128 binary inputs to maintain semantic meaning. The look up table therefore gets large, causing the agent to soak compute thus slowing lookup calculations. The other option is to have a set of genres with known embeddings and a binary encoding for each. Every time a new genre comes in, a distance calculation could work out which "bucket" it would fall under and use that binary input.

## Embedding bucketing Example

### Embedding Cache

| Buckets | Embeddings | Binary Encoding |
|---|---|---|
| Drama | 0.2343, -0.29323... | 0,0 |
| Horror | 0.9342, -0.24326... | 0,1 |
| Romance | 0.1392, -0.04322... | 1, 1 |

There is an input genre of action and we get its embedding

| Input | Action |
|---|---|
| embedding | 0.3849, -0.4932 |

Next, we do a cosine similarity calculation to find the closest bucket.

| Buckets | Distance |
|---|---|
| drama | 0.2 |
| horror | 0.5 |
| romance | 0.8 |

Pick the closest one, in this case, *drama* and use its encoding 0,0. In this way, we are able to account for an unknown input size space.

WNNs are notorious regarding scalability [4]. The model's approach relies on lookup operations for every training and inference event. With large lookup tables when the agent has lots of learning events, the WNN struggles with compute. This is also illustrated on the table of results under the quantitative analysis section. The WNN takes 10x the time. It is important to note, however, that the agents here are running on CPU and have not been optimized for GPU usage yet. Initial tests using PyTorch to facilitate the GPU offload operations indicate promising speed increases in terms of both training and testing time.

## Future work

This novel approach has not yet been fully explored but future work would include optimisations of the WNN for compute, running on GPUs via PyTorch and experimenting with more hidden Q layers to capture more complicated patterns.We also briefly explore combinations of both weighted and weightless networks in a combined system to capture both the versatility of current neural networks and the data efficiency of WNNs.

## Conclusion

We explored the hypothesis that smaller, personal datasets when applied using the right algorithm can beat out traditional larger models trained on massed data, suggesting potential with weightless neural networks. These advantages also help with the cold-start problem and suggest a way towards more controllable, interpretable, and efficient recommendation algorithms.